\documentclass[twocolumn,showpacs,superscriptaddress,longbibliography,prl]{revtex4-1}

\usepackage{graphicx}
\usepackage{dcolumn}
\usepackage{amsmath}
\usepackage{color}
\usepackage{braket}
\usepackage{float}
\usepackage[english]{babel}

\begin{document}

\title{Magnetic Excitations and Continuum of a Field-Induced Quantum Spin Liquid in $\alpha$-RuCl$_3$}

\author{Zhe~Wang}
\affiliation{Institute of Radiation Physics, Helmholtz-Zentrum Dresden-Rossendorf, 01328 Dresden, Germany}

\author{S. Reschke}
\affiliation{Experimental Physics V, Center for Electronic
Correlations and Magnetism, Institute of Physics, University of Augsburg, 86135 Augsburg, Germany}

\author{D. H\"{u}vonen}
\affiliation{National Institute of Chemical Physics and Biophysics, Akadeemia tee 23, 12618 Tallinn, Estonia}

\author{S.-H.~Do}
\author{K.-Y.~Choi}
\affiliation{Department of Physics, Chung-Ang University, Seoul 06974, Republic of Korea}

\author{M.~Gensch}
\affiliation{Institute of Radiation Physics, Helmholtz-Zentrum Dresden-Rossendorf, 01328 Dresden, Germany}

\author{U. Nagel}
\author{T. R\~{o}\~{o}m}
\affiliation{National Institute of Chemical Physics and Biophysics, Akadeemia tee 23, 12618 Tallinn, Estonia}

\author{A.~Loidl}
\affiliation{Experimental Physics V, Center for Electronic
Correlations and Magnetism, Institute of Physics, University of Augsburg, 86135 Augsburg, Germany}

\date{\today}

\begin{abstract}
We report on terahertz spectroscopy of quantum spin dynamics in $\alpha$-RuCl$_3$,
a system proximate to the Kitaev honeycomb model,
as a function of temperature and magnetic field.
An extended magnetic continuum develops
below the structural phase transition at $T_{s2}=62$~K.
With the onset of a long-range magnetic order at $T_N=6.5$~K,
spectral weight is transferred to a well-defined magnetic excitation at $\hbar \omega_1 = 2.48$~meV,
which is accompanied by a higher-energy band at $\hbar \omega_2 = 6.48$~meV.
Both excitations soften in magnetic field, signaling a quantum phase transition at $B_c=7$~T
where we find a broad continuum dominating the dynamical response.
Above $B_c$, the long-range order is suppressed, and on top of the continuum, various emergent magnetic excitations evolve.
These excitations follow clear selection rules and exhibit distinct field dependencies,
characterizing the dynamical properties of the field-induced quantum spin liquid.
\end{abstract}

\maketitle

Quantum spin liquids (QSLs) are exotic states of matter
in which quantum fluctuations prevent conventional magnetic long-range order even at the lowest temperatures.
The ground states of QSLs are highly-entangled and albeit disordered,
can exhibit well-defined quasiparticles which are non-local and fractionalized \cite{Balents17,Balents10}.
The Kitaev honeycomb model \cite{Kitaev06} is a representative example for the existence of a QSL,
which hosts fractionalized Majorana fermions and flux excitations.
Moreover, bound states of the fractional excitations and excitations of non-Abelian type can occur
when the time-reversal symmetry is broken \cite{Trebst14,Knolle15,ThVojta17},
e.g, by applying an external magnetic field.
Also, as this model is exactly solvable and thus provides quantitative understanding
for the ground state as well as for the dynamical response of the QSL \cite{Kitaev06,Knolle14},
significant experimental efforts have been motivated to realize the QSL and to search for its unconventional excitations.

Strong spin-orbit couplings together with a proper crystal electric field are important ingredients for realizing the Kitaev honeycomb model in magnetic insulators \cite{Jackeli09}.
Based on $5d$ or $4d$ ions with strong spin-orbit coupling, a Mott insulator with an effective spin-1/2 localized on a honeycomb lattice has been realized in the iridates \cite{Singh12} and recently in $\alpha$-RuCl$_3$ \cite{Plumb14,Kubota15,Johnson15,Cao16,Park16}.
In $\alpha$-RuCl$_3$, besides the significant Kitaev term \cite{Sandilands15,Do17} that accounts for the bond-dependent Ising exchanges,
other interactions, especially the Heisenberg and off-diagonal exchange interactions,
have been suggested to be responsible for the magnetic zigzag order below $T_N=6.5$~K and for the observed sharp spin-wave-like excitations \cite{Banerjee16NM,Banerjee16,Ran17,Winter16,Winter17,Janssen16}.

\begin{figure}[t]
\centering
\includegraphics[width=75mm,clip]{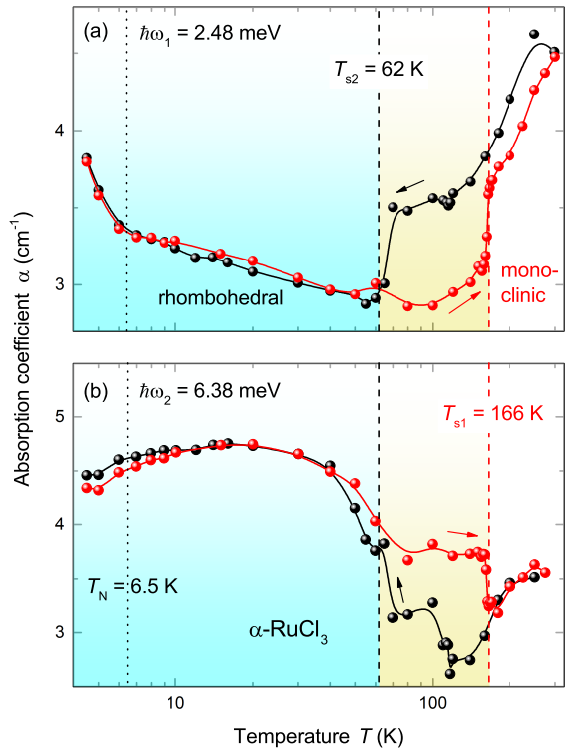}
\vspace{2mm} \caption[]{\label{Fig:Tdep_Abs}
Absorption coefficient $\alpha$ as a function of temperature $T$ for the photon energies (a) $\hbar \omega_1 = 2.48$~meV and (b) $\hbar \omega_2 = 6.38$~meV, respectively, measured with the THz wave vector perpendicular to the crystallographic $ab$ plane of $\alpha$-RuCl$_3$. Clear hysteresis is revealed at the structural phase transitions between the low-temperature rhombohedral and the high-temperature monoclinic phases.
Solid lines are guides for the eyes.
The structural phase boundaries are indicated by the dashed lines at $T_{s2}=62$~K and $T_{s1}=166$~K,
and the magnetic phase transition by the dotted line at $T_N=6.5$~K.
}
\end{figure}

To realize a QSL in $\alpha$-RuCl$_3$, obviously the long-range zigzag order has to be suppressed, e.g. by tuning the external parameters, such as pressure \cite{LilingSun17,WeiqiangYu17} or magnetic field \cite{Baek17,Hentrich17,Sears17,Wolter17,Zheng17,Kubota15}.
When the external magnetic field is applied in the crystallographic $ab$ plane of $\alpha$-RuCl$_3$, the magnetic zigzag order is suppressed at about $B_c=7$~T, with the emergence of a field-induced disordered phase, a putative QSL.
However, whether the elementary spin excitations are gapped or gapless, which is important for identifying a QSL \cite{Balents17}, is still under debate for the field-induced disordered phase in $\alpha$-RuCl$_3$ \cite{Baek17,Zheng17},
since a direct probe of the magnetic excitations in the disordered phase is missing.
Moreover, the existence of exotic quasiparticles is yet to be proven for the time-reversal symmetry-broken phase in the applied magnetic field.

Here, by performing terahertz (THz) spectroscopy as a function of temperature and magnetic field,
we are able to directly reveal the quantum spin dynamics in the different phases of $\alpha$-RuCl$_3$.
We resolve a magnetic continuum at zero field, which develops below the structural phase transition at $T_{s2}=62$~K.
Across the magnetic ordering temperature $T_N = 6.5$~K, the spectral weight of the continuum is partially shifted to two magnetic excitations.
As a function of magnetic field applied in the crystallographic $ab$ plane,
we provide spectroscopic evidence for a field-induced quantum phase transition at $B_c=7$~T.
Approaching $B_c$ from below, the two magnetic excitations soften and we observe a remarkably broad continuum just at the critical field.
Above $B_c$, various magnetic excitations show up signaling the opening of a spin gap.
The hierarchy and distinct field dependencies of the emergent excitations characterize the dynamical properties,
which could be viewed as a signature of many-body interactions and/or bound states of fractionalized excitations \cite{Trebst14,Knolle15,ThVojta17} in the field-induced QSL.
Our experimental results also reveal a clear polarization dependence of these excitations,
establishing a comprehensive characterization of the quantum spin dynamics of the QSL.


High-quality single crystals of $\alpha$-RuCl$_3$ were grown using a vacuum sublimation method \cite{Do17} and have been characterized by various techniques \cite{Park16,Wolter17,Reschke17,Choi17}.
The samples for the optical experiments have a typical $ab$ surface of $5\times3$~mm$^2$ and a thickness of 1~mm.
Time-domain THz transmission measurements were performed with the THz wave vector $\mathbf{k}$ perpendicular to the crystallographic $ab$ plane,
for 4--300~K in the spectral range 0.1--3.3~THz using a TPS Spectra 3000 spectrometer (TeraView Ltd.).
Time domain signals were obtained for references (empty apertures) and samples,
from which the power spectra could be evaluated via Fourier transformation.
Field dependent THz absorption experiments  were carried out in Voigt configuration,
i.e. $\mathbf{k} \perp \mathbf{B}$,
with a magneto-optic cryostat equipped with a 17~T superconducting magnet.
The absorption spectra were recorded using a Sciencetech SPS200 Martin-Puplett type spectrometer
with a 0.3 K bolometer and a rotatable polarizer in front of the sample.

\begin{figure}[t]
\centering
\includegraphics[width=75mm,clip]{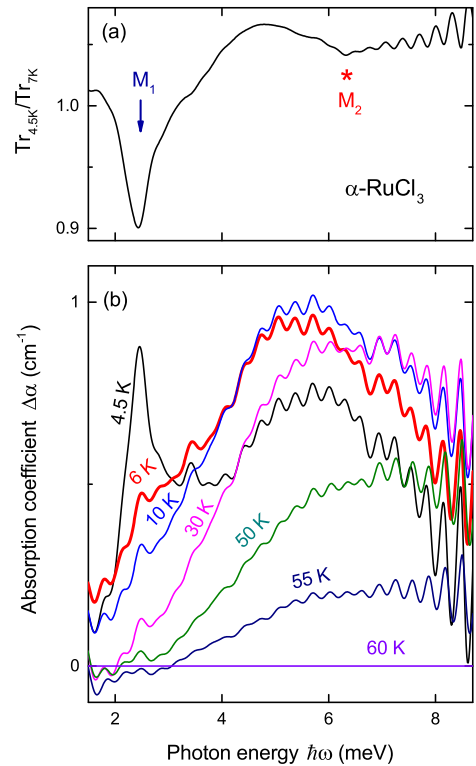}
\vspace{2mm} \caption[]{\label{Fig:freqdep}
(a) Ratio of transmission as obtained at 4.5~K ($T < T_N$) and 7~K ($T > T_N$) as a function of photon energy in $\alpha$-RuCl$_3$.
Two dips appear below $T_N=6.5$~K at $\hbar \omega_1 = 2.48$~meV (mode M$_1$) and $\hbar \omega_2 = 6.38$~meV (band M$_2$), as marked by the arrow and asterisk, respectively.
(b) Evolution of absorption-coefficient spectra with decreasing temperature below $T_{s2}$, where the 60~K spectrum is taken as a reference.
A broad continuum extending over the whole spectral range evolves with decreasing temperature down to $T_N$.
Below $T_N$, the spectral weight transfers to lower frequency with a sharp peak developed at $\hbar \omega_1$.
The oscillations are due to multiple interference at the sample surfaces.
}
\end{figure}

THz response of $\alpha$-RuCl$_3$ exhibits a very anomalous temperature dependence.
Figure~\ref{Fig:Tdep_Abs} shows the absorption coefficient as a function of temperature measured by heating and cooling the sample, for the photon energies $\hbar \omega_1 = 2.48$~meV and $\hbar \omega_2 = 6.38$~meV [see Fig.~\ref{Fig:freqdep}(a)].
For both frequencies we observe a markedly strong hysteresis between the heating and cooling curves for the temperature range from $T_{s2}=62$~K to $T_{s1}=166$~K.
This confirms the structural transition between the high-temperature monoclinic and the low-temperature rhombohedral phases \cite{Park16,Baek17,Choi17,Reschke17},
and also indicates that above $T_{s2}$ the THz response mainly reflects strong structural fluctuations.

THz spectroscopy is sensitive to dynamical response in the vicinity of the $\Gamma$ point,
where inelastic neutron scattering experiments have revealed a broad continuum of magnetic quasiparticles up to 120~K \cite{Banerjee16,Do17}.
The evolution of this continuum with temperature was compared to an expected temperature dependence of the continuum of fractional Majorana fermions in the Kitaev model \cite{Do17,Yoshitake16}.
To avoid the complication of the structural fluctuations above $T_{s2}$,
we show in Fig.~\ref{Fig:freqdep}(b) the absorption spectra below $T_{s2}$ with the spectrum of 60~K taken as a reference.
It has been indicated by other experimental techniques that below this temperature,
an evident magnetic-field dependent behavior develops \cite{Baek17,Hentrich17}.
Compared to the inelastic neutron scattering results, the THz spectroscopy reveals here a similar broad continuum extending over the spectral range up to 9~meV.
With decreasing temperature the continuum develops rapidly and exhibits a broad maximum with the maximum position shifting to lower frequency.
At 6~K (below $T_N$), the continuum is slightly reduced and the spectral weight is transferred to an excitation appearing at about 2.5~meV.
The transfer of spectral weight to lower energies becomes more evident at lower temperatures.
As seen from the 4.5~K spectrum, a very sharp peak shows up at 2.48~meV, while the continuum with maximum at 5--6~meV is strongly suppressed.
This behavior is also clearly displayed in Fig.~\ref{Fig:Tdep_Abs}.
While the absorption coefficient at 2.48~meV increases below $T_N$, it drops slightly around 6.38~meV.

To unambiguously identify the magnetic excitations, we calculate the ratio of transmission as measured at 4.5 and 7~K, below and right above the magnetic ordering temperature.
As shown in Fig.~\ref{Fig:freqdep}(a), the ratio of transmission exhibits a sharp dip at 2.48~meV (mode M$_1$) and a shallow minimum around 6.38~meV (band M$_2$).
Both modes, with frequencies comparable to the magnetic excitations determined in the inelastic neutron scattering experiments \cite{Banerjee16,Do17},
have been captured by recent exact-diagnolization calculations \cite{Winter17}
of a model guided by \emph{ab-initio} studies \cite{Winter16,Yadav16,Li16,Xiang16},
where in addition to the Kitaev interactions,
the off-diagonal couplings are included together with the nearest- and third-neighbour Heisenberg exchange interactions.
However, the nature of the two modes remains unsettled.
The lower-lying mode M$_1$ with a sharp absorption line, ascribed to the antiferromagnetic resonance in Ref.~\cite{Little17},
corresponds to the single-magnon branch at the $\Gamma$ point in the calculations \cite{Winter17}.
In contrast, the band M$_2$ at higher energies is rather broad and weak.
This mode may correspond to the higher-energy mode with small intensity in the linear spin-wave theory \cite{Winter17}.
Since M$_1$ and M$_2$ arise below the antiferromagnetic phase transition,
they should characterize the dynamical response of the zig-zag order below $T_N$.
Above $T_N$, the modes M$_1$ and M$_2$ transfer their spectral weight to the continuum as observed in the THz spectra [Fig.~\ref{Fig:freqdep}(b)].
The temperature dependence of the continuum is consistent with the results of the neutron scattering study \cite{Do17,Banerjee16},
which has been discussed in the context of the Kitaev honeycomb model \cite{Yoshitake16,Banerjee16}.
Alternatively, since the non-Kitaev terms are also important for $\alpha$-RuCl$_3$ \cite{Ran17,Winter17}, one may speculate that this experimentally resolved continuum could be a two-magnon continuum, which can be excited due to the anharmonic terms of the magnon interactions \cite{Winter17}.

\begin{figure*}[t]
\centering
\includegraphics[width=175mm,clip]{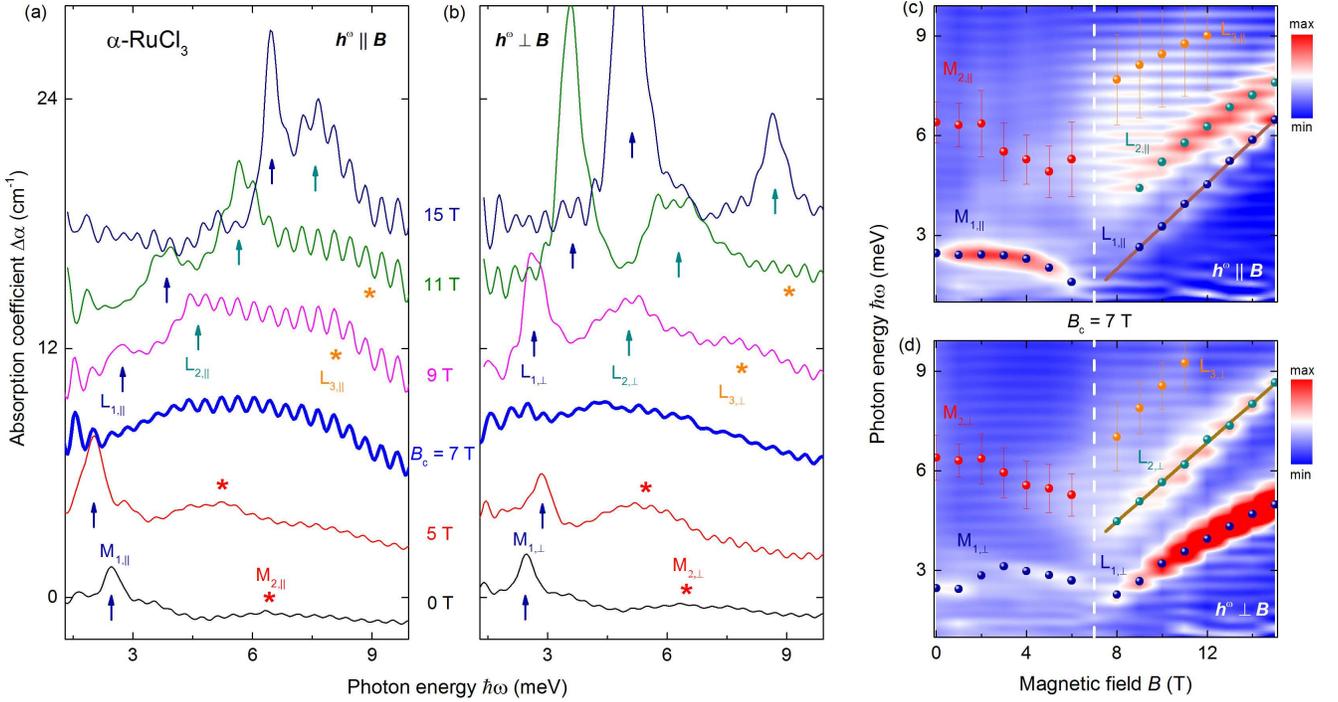}
\vspace{2mm} \caption[]{\label{Fig:Fdep}
THz absorption spectra of $\alpha$-RuCl$_3$ measured at 2.4~K below and above the critical field $B_c=7$~T for (a) the THz magnetic field parallel ($\mathbf{h}^\omega \parallel \mathbf{B}$) and (b) perpendicular to the external magnetic field ($\mathbf{h}^\omega \perp \mathbf{B}$), respectively,
with the spectra of 10~K taken as references.
The spectra are shifted vertically by a constant for clarity.
The arrows indicate the low-energy magnetic excitations with sharp absorption lines, while the asterisks mark the high-energy excitations that are rather broad.
At the critical field, a broad continuum is observed which extends over the whole spectral range.
Above $B_c$, various excitations emerge on the top of the continuum.
(c)(d) Contour plot of the absorption coefficient as a function of field and photon energy for $\mathbf{h}^\omega \parallel \mathbf{B}$ and $\mathbf{h}^\omega \perp \mathbf{B}$.
The symbols correspond to the peak positions of the low-energy modes and the maximum positions of the high-energy bands in (a) and (b).
For the broad bands, the linewidths are indicated by the error bars.
Above $B_c$, for $\mathbf{h}^\omega \parallel \mathbf{B}$ the lower-lying mode L$_{1,\parallel}$ follows a linear field dependence with an apparent $g$ value of $g^\ast=11.1(1)$ assuming a magnetic-dipole excitation [solid line in (c)], while for $\mathbf{h}^\omega \perp \mathbf{B}$ the field dependence of the higher-lying mode L$_{2,\perp}$ is described by the linear Zeeman term with $g^\ast=10.2(2)$ [solid line in (d)].
The vertical dashed line marks the critical field $B_c=7$~T.
}
\end{figure*}

Figures~\ref{Fig:Fdep}(a) and~\ref{Fig:Fdep}(b) show the absorption  spectra as a function of an external magnetic field applied in the $ab$ plane, for the THz magnetic field $\mathbf{h}^\omega \parallel \mathbf{B}$ and $\mathbf{h}^\omega \perp \mathbf{B}$, respectively. The spectra at 10~K in zero field are taken as the respective reference.
For $\mathbf{h}^\omega \parallel \mathbf{B}$, the lower-lying mode M$_{1,\parallel}$ softens and becomes sharper in finite fields, together with the softening of the broad band M$_{2,\parallel}$ [see Fig.~\ref{Fig:Fdep}(c)].
At the critical field $B_c=7$~T, the dynamical response is dominated by a broad continuum, extending over the whole resolvable spectra range
(with the oscillations resulting from multiple interference at the sample surfaces).
On further increasing field, the broad continuum shifts to higher energies with three new excitations appearing.
At 9~T, the three excitations are observed at 2.7, 4.4, and 8.1~meV, denoted by L$_{1,\parallel}$, L$_{2,\parallel}$, and  L$_{3,\parallel}$, respectively, and they can be well tracked and shift to higher energies with increasing magnetic field.
While the two lower-lying modes L$_{1,\parallel}$ and L$_{2,\parallel}$ are well-defined
and become narrower in higher fields,
the high-energy excitation L$_{3,\parallel}$ is a broad band and shifts out of the spectral range above 12~T.
Due to the broad-band nature and small intensity, L$_{3,\parallel}$ cannot be fully resolved,
thus it could also be the higher-energy flank of the continuum, which shifts to higher energies in the magnetic fields.

The field dependence of these modes and the emergent continuum is represented in the contour plot of Fig.~\ref{Fig:Fdep}(c).
The softening feature close to the critical field indicates that the spin gap is closed and another one is reopened, in agreement with the results obtained from other methods \cite{Sears17,Baek17,Hentrich17}.
A very interesting feature of the field-induced disordered phase is the linear field dependence of the lower-lying mode L$_{1,\parallel}$.
As shown in Fig.~\ref{Fig:Fdep}(c) this mode follows $\hbar \omega = \hbar \omega_0 + g^\ast\mu_BB \Delta S$ with
an apparent $g$-value of $g^\ast=11.1(1)$ and the Bohr magneton $\mu_B$, assuming a magnetic-dipole transition, i.e. $\Delta S=1$.
This suggests that a spin gap opens linearly with the magnetic field for the sector of the dynamical structure factor corresponding to $\mathbf{h}^\omega \parallel \mathbf{B}$.
The linear opening of the spin gap is a feature expected for models beyond the Kitaev limit \cite{Song16}.
The apparent $g^\ast$-factor revealed here is significantly larger than the value of $g_{ab}= 2.5$ estimated from the magnetization measurements \cite{Kubota15} or $g_{ab}= 2.27$ from the x-ray magnetic circular dichroism experiments \cite{Agrestini17},
which is a reminiscence of the existence of bound states and many-body interactions in the QSL, as for the Heisenberg spin-1/2 chains \cite{WangBethe,Yang17}.

For $\mathbf{h}^\omega \perp \mathbf{B}$ [Fig.~\ref{Fig:Fdep}(b)], approaching the critical field from below, the lower-lying mode M$_{1,\perp}$ first slightly hardens and then softens, while the band M$_{2,\perp}$ softens monotonically.
At the critical field, the spectrum is again dominated by a broad continuum extending the whole spectral range.
Above $B_c$ one can identify three modes: the two low-lying sharp modes L$_{1,\perp}$ and L$_{2,\perp}$, and the higher-energy band L$_{3,\perp}$, emerging on top of the continuum and shifting to higher energy together with the continuum.
Both of the two lower-lying modes become sharper with increasing field,
while the band L$_{3,\perp}$ remains broad and thus could merely be a feature of the continuum shifting in the magnetic field,
similar as for L$_{3,\parallel}$.
In addition, the continuum is suppressed in high fields and the spectral weight is transferred to the low-lying modes.
In the contour plot of Fig.~\ref{Fig:Fdep}(d) one can readily see these field-dependent features.

The emergent magnetic excitations and continuum of the field-induced QSL exhibit clear contrast between
$\mathbf{h}^\omega \perp \mathbf{B}$ and $\mathbf{h}^\omega \parallel \mathbf{B}$.
Both the eigenenergies and their field dependencies are different for the two polarizations,
revealing the contrast of the two different sectors of dynamical structure factors.
For $\mathbf{h}^\omega \perp \mathbf{B}$ the lower-lying mode L$_{1,\perp}$ has lower eigenenergies, and the absorption lines are much sharper and stronger than those of its counterpart L$_{1,\parallel}$.
Moreover, the eigenenergy of L$_{1,\perp}$ does not follow a linear field dependence, but a slower increase with the increasing field.
The slower increase is consistent with the results of low-temperature specific heat measurements (below 3~K) \cite{Wolter17}
where the thermodynamics is presumably dominated by the lowest-lying excitations.
In contrast, the mode L$_{2,\perp}$ in the sector of $\mathbf{h}^\omega \perp \mathbf{B}$ can be described by the linear Zeeman term with $g^\ast=10.2(2)$, while the L$_{2,\parallel}$ mode hardens evidently slower than linear with the field, although these two modes have similar linewidths.
These modes could be the bound states of fractionalized excitations \cite{Knolle15,Trebst14,ThVojta17},
which are narrow and can evolve from the excitation continuum as the time-reversal symmetry is broken by the applied magnetic field \cite{KnolleThesis}.
For the higher-energy bands L$_{3,\parallel}$ and L$_{3,\perp}$, the field dependencies are also sensitive to the polarizations.
Together with the underlying continuum or as a part of the continuum, their spectral weights are reduced in high magnetic fields.

To conclude, terahertz spectroscopy as a function of temperature and magnetic field
reveals emergent magnetic excitations and continua in a field-induced QSL in $\alpha$-RuCl$_3$.
While two magnonic excitations evolving on top of a continuum are resolved in the magnetically ordered phase and well below the critical field,
the QSL is dynamically characterized by an emergent continuum at the quantum criticality
and various magnetic excitations with distinct dependencies on field and polarization above the quantum phase transition.
Our results pave the way to understand the dynamical behavior of the QSL
and provide constrains for a quantitative theoretical description,
e.g. via the identification of a realistic model or by extending the Kitaev model with terms breaking the time-reversal symmetry.

\begin{acknowledgements}
We would like to thank  Seung-Ho Baek,  Lukas Janssen, Johannes Knolle, Roderich Moessner, Roser Valent\'{\i}, Jinsheng Wen, Stephen M. Winter, Liang Wu, Yi-Zhuang You, and  Xenophon Zotos for the helpful discussions.
This work has been partially supported by the Deutsche Forschungsgemeinschaft via the Transregional Research Collaboration TRR 80: From Electronic Correlations to Functionality (Augsburg - Munich - Stuttgart), and a Korea Research Foundation (KRF) Grant, funded by the Korean Government (MEST) (Grant No.: 20160874). The work in Tallinn was supported by the Estonian Ministry of Education and Research under Grant No. IUT23-03, the Estonian Research Council Grant No. PUT451, and the European Regional Development Fund project TK134.
\end{acknowledgements}


\begin{thebibliography}{39}

\bibitem{Balents17}
L. Savary and L. Balents,
Quantum spin liquids: a review.
Rep. Prog. Phys. \textbf{80}, 016502 (2017).

\bibitem{Balents10}
L. Balents,
Spin liquids in frustrated magnets.
Nature \textbf{464}, 199 (2010).

\bibitem{Kitaev06}
A. Kitaev,
Anyons in an exactly solved model and beyond.
Ann. Phys. \textbf{321} 2–11 (2006).

\bibitem{Knolle15}
J. Knolle, D. L. Kovrizhin, J. T. Chalker, and R. Moessner,
Dynamics of fractionalization in quantum spin liquids.
Phys. Rev. B \textbf{92}, 115127 (2015).

\bibitem{Trebst14}
V. Lahtinen, A. W. W. Ludwig, and S. Trebst,
Perturbed vortex lattices and the stability of nucleated topological phases.
Phys. Rev. B \textbf{89}, 085121 (2014).

\bibitem{ThVojta17}
H. Th\'{e}veniaut and M. Vojta,
Bound states of fractionalized excitations in a modulated Kitaev spin liquid.
arXiv:1705.08913 (2017).

\bibitem{Knolle14}
J. Knolle,  D. L. Kovrizhin,  J. T. Chalker, and  R. Moessner,
Dynamics of a Two-Dimensional Quantum Spin Liquid: Signatures of Emergent Majorana Fermions and Fluxes
Phys. Rev. Lett. \textbf{112}, 207203 (2014).



\bibitem{Jackeli09}
G. Jackeli and G. Khaliullin,
Mott Insulators in the Strong Spin-Orbit Coupling Limit:
From Heisenberg to a Quantum Compass and Kitaev Models.
Phys. Rev. Lett. \textbf{102}, 017205 (2009).

\bibitem{Singh12}
Y. Singh, S. Manni, J. Reuther, T. Berlijn, R. Thomale, W. Ku, S. Trebst, and P. Gegenwart,
Relevance of the Heisenberg-Kitaev Model for the Honeycomb Lattice Iridates $A_2$IrO$_3$
Phys. Rev. Lett. \textbf{108}, 127203 (2012).


\bibitem{Plumb14}
K. W. Plumb, J. P. Clancy, L. J. Sandilands, V. Vijay Shankar, Y. F. Hu, K. S. Burch, Hae-Young Kee, and Young-June Kim,
$\alpha$-RuCl$_3$: A spin-orbit assisted Mott insulator on a honeycomb lattice.
Phys. Rev. B \textbf{90}, 041112(R) (2014).

\bibitem{Kubota15}
Y. Kubota, H. Tanaka, T. Ono, Y. Narumi, and K. Kindo,
Successive magnetic phase transitions in $\alpha$-RuCl$_3$: XY-like frustrated magnet on the honeycomb lattice.
Phys. Rev. B \textbf{91}, 094422 (2015).

\bibitem{Johnson15}
R. D. Johnson, S. C. Williams, A. A. Haghighirad, J. Singleton, V. Zapf, P. Manuel, I. I. Mazin, Y. Li, H. O. Jeschke, R. Valentí, and R. Coldea,
Monoclinic crystal structure of $\alpha$-RuCl$_3$ and the zigzag antiferromagnetic ground state.
Phys. Rev. B \textbf{92}, 235119 (2015).

\bibitem{Cao16}
H. B. Cao, A. Banerjee, J.-Q. Yan, C. A. Bridges, M. D. Lumsden, D. G. Mandrus, D. A. Tennant,
B. C. Chakoumakos, and S. E. Nagler,
Low-temperature crystal and magnetic structure of $\alpha$-RuCl$_3$.
Phys. Rev. B \textbf{93}, 134423 (2016).

\bibitem{Park16}
S.-Y. Park, S.-H. Do, K.-Y. Choi, D. Jang, T.-H. Jang, J. Schefer, C.-M. Wu, J. S. Gardner, J.
M. S. Park, J.-H. Park, and Sungdae Ji,
Emergence of the Isotropic Kitaev Honeycomb Lattice with Two-dimensional Ising Universality in $\alpha$-RuCl$_3$.
arXiv:1609.05690 (2016).



\bibitem{Sandilands15}
L. J. Sandilands, Y. Tian, K. W. Plumb, Y.-J. Kim, and K. S. Burch,
Scattering Continuum and Possible Fractionalized Excitations in $\alpha$-RuCl$_3$.
Phys. Rev. Lett. \textbf{114}, 147201 (2015).

\bibitem{Do17}
S.-H. Do, S.-Y. Park, J. Yoshitake, J. Nasu, Y. Motome, Y.
S. Kwon, D. T. Adroja, D. J. Voneshen, K. Kim, T.-H. Jang, J.-H. Park,
K.-Y. Choi, and S. Ji,
Incarnation of Majorana Fermions in Kitaev Quantum Spin Lattice.
arXiv:1703.01081 (2017).

\bibitem{Banerjee16NM}
A. Banerjee, C. A. Bridges, J.-Q. Yan, A. A. Aczel, L. Li, M. B. Stone, G. E. Granroth,
M. D. Lumsden, Y. Yiu, J. Knolle, S. Bhattacharjee, D. L. Kovrizhin, R. Moessner, D. A. Tennant,
D. G. Mandrus, and S. E. Nagler,
Proximate Kitaev quantum spin liquid behaviour in a honeycomb magnet.
Nature Materials \textbf{15}, 733–740 (2016).

\bibitem{Ran17}
K. Ran,  J.i Wang, W. Wang, Z.-Y. Dong, X. Ren, S. Bao, S. Li, Z. Ma, Y. Gan,
Y. Zhang, J. T. Park, G. Deng, S. Danilkin, S.-L. Yu, J.-X. Li, and J. Wen,
Spin-Wave Excitations Evidencing the Kitaev Interaction in Single Crystalline $\alpha$-RuCl$_3$.
Phys. Rev. Lett. \textbf{118}, 107203 (2017).

\bibitem{Banerjee16}
A. Banerjee, J. Yan, J. Knolle, C. A. Bridges, M. B. Stone, M. D. Lumsden, D. G. Mandrus, D. A. Tennant, R. Moessner, and S. E. Nagler,
Neutron tomography of magnetic Majorana fermions in a proximate quantum spin liquid.
arXiv:1609.00103 (2016).

\bibitem{Winter17}
S. M. Winter,   K. Riedl, A. Honecker, and R. Valent\'{\i},
Breakdown of Magnons in a Strongly Spin-Orbital Coupled Magnet.
arXiv:1702.08466 (2017).

\bibitem{Winter16}
S. M. Winter,  Y. Li,   H. O. Jeschke, and  R. Valent\'{\i},
Challenges in design of Kitaev materials: Magnetic interactions from competing energy scales.
Phys. Rev. B \textbf{93}, 214431 (2016).

\bibitem{Janssen16}
L. Janssen,  E. C.  Andrade, and  M. Vojta,
Honeycomb-Lattice Heisenberg-Kitaev Model in a Magnetic Field:
Spin Canting, Metamagnetism, and Vortex Crystals.
Phys. Rev. Lett. \textbf{117}, 277202 (2016).

\bibitem{LilingSun17}
Z. Wang, J. Guo, F. F. Tafti, A. Hegg, S. Sen, V. A. Sidorov,
L. Wang, S. Cai, W. Yi, Y. Zhou, H. Wang, S. Zhang, K. Yang,
A. Li, X. Li, Y. Li, J. Liu, Y. Shi, W. Ku, Q. Wu,
R. J. Cava, and L. Sun,
Observation of the quantum spin liquid state in pressurized $\alpha$-RuCl$_3$.
arXiv:1705.06139 (2017).

\bibitem{WeiqiangYu17}
Y. Cui, J. Zheng, K. Ran, J. Wen, Z. Liu, B. Liu, W. Guo, and W. Yu,
A pressure-induced quantum disordered phase in $\alpha$-RuCl$_3$ evidenced by NMR.
arXiv:1706.02697 (2017).

\bibitem{Wolter17}
A.U.B. Wolter, L. T. Corredor, L. Janssen, K. Nenkov, S. Sch\"{o}necker, S.-H.
Do, K.-Y. Choi, R. Albrecht, J. Hunger, T. Doert, M. Vojta, and B. B\"{u}chner,
Field-induced quantum criticality in the Kitaev system $\alpha$-RuCl$_3$.
arXiv:1704.03475 (2017).


\bibitem{Sears17}
J. A. Sears, Y. Zhao, Z. Xu, J. W. Lynn, and  Y.-J. Kim,
Phase Diagram of $\alpha$-RuCl$_3$ in an in-plane Magnetic Field.
arXiv:1703.08431 (2017).


\bibitem{Baek17}
S.-H. Baek, S.-H. Do, K.-Y. Choi, Y. S. Kwon, A.U.B.
Wolter, S. Nishimoto, J. van den Brink, and B. B\"{u}chner,
Observation of a Field-induced Quantum Spin Liquid in $\alpha$-RuCl$_3$.
arXiv:1702.01671 (2017).


\bibitem{Zheng17}
J. Zheng, K. Ran, T. Li, J. Wang, P. Wang,
B. Liu, Z. Liu, B. Normand, J. Wen, and Weiqiang Yu,
Gapless Spin Excitations in the Field-Induced Quantum Spin Liquid Phase of $\alpha$-RuCl$_3$.
arXiv:1703.08474 (2017).

\bibitem{Hentrich17}
R. Hentrich, A.U.B. Wolter, X. Zotos, W. Brenig, D. Nowak,
A. Isaeva, T. Doert, A. Banerjee, P. Lampen-Kelley, D. G. Mandrus,
S. E. Nagler, J. Sears, Y.-J. Kim, Bernd B\"{u}chner, and Ch. Hess,
Large field-induced gap of Kitaev-Heisenberg paramagnons in $\alpha$-RuCl$_3$.
arXiv:1703.08623 (2017).





\bibitem{Choi17}
A. Glamazda, P. Lemmens, S.-H. Do, Y. S. Kwon, and K.-Y. Choi,
Relation between Kitaev magnetism and structure in $\alpha$-RuCl$_3$
Phys. Rev. B \textbf{95}, 174429 (2017).

\bibitem{Reschke17}
S. Reschke, F. Mayr, Zhe Wang, Seung-Hwan Do, K.-Y. Choi, A. Loidl,
Electronic and phonon excitations in $\alpha$-RuCl$_3$.
arXiv:1706.02724 (2017).

\bibitem{Yoshitake16}
J. Yoshitake,  J. Nasu, and Y. Motome,
Fractional Spin Fluctuations as a Precursor of Quantum Spin Liquids:
Majorana Dynamical Mean-Field Study for the Kitaev Model.
Phys. Rev. Lett. \textbf{117}, 157203 (2016).


\bibitem{Yadav16}
R. Yadav, N. A. Bogdanov, V. M. Katukuri, S. Nishimoto,
J. van den Brink, and L. Hozoi,
Kitaev exchange and field-induced
quantum spin-liquid states in
honeycomb $\alpha$-RuCl$_3$.
Sci. Rep. \textbf{6}, 37925 (2016).

\bibitem{Li16}
W. Wang, Z.-Y. Dong, S.-L. Yu, and J.-X. Li,
Theoretical investigation of the magnetic dynamics and superconducting pairing symmetry in $\alpha$-RuCl$_3$.
arXiv:1612.09515 (2016).

\bibitem{Xiang16}
Y. S. Hou, H. J. Xiang, and X. G. Gong,
Unveiling Magnetic Interactions of Ruthenium Trichloride via Constraining
Direction of Orbital moments: Potential Routes to Realize Quantum Spin Liquid.
arXiv:1612.00761 (2016).

\bibitem{Little17}
A. Little, Liang Wu, P. Lampen-Kelley, A. Banerjee, S. Pantankar, D. Rees,
 C. A. Bridges, J.-Q. Yan, D. Mandrus, S. E. Nagler, and J. Orenstein,
Antiferromagnetic resonance and terahertz conductivity in $\alpha$-RuCl$_3$.
arXiv:1704.07357 (2017).

\bibitem{Song16}
X.-Y. Song, Y.-Z. You, and L. Balents,
Low-Energy Spin Dynamics of the Honeycomb Spin Liquid Beyond the Kitaev Limit.
Phys. Rev. Lett. 117, 037209 (2016).

\bibitem{Agrestini17}
S. Agrestini, C.-Y. Kuo, K.-T. Ko, Z. Hu, D. Kasinathan, H. Babu Vasili, J. Herrero-Martin, S. M. Valvidares,
E. Pellegrin, L.-Y. Jang, A. Henschel, M. Schmidt, A. Tanaka, and L. H. Tjeng,
Electronically highly cubic conditions for Ru in $\alpha$-RuCl$_3$.
arXiv:1704.05100 (2017).

\bibitem{WangBethe}
Zhe Wang, Jianda Wu, Wang Yang, Anup Kumar Bera, Dmytro Kamenskyi, A.T.M. Nazmul Islam, Shenglong Xu, Joseph Matthew Law, Bella Lake, Congjun Wu, Alois Loidl,
Experimental Observation of Bethe Strings.
arXiv:1706.04181 (2017).

\bibitem{Yang17}
W. Yang, J. Wu, S. Xu, Z. Wang, C. Wu,
Quantum spin dynamics of the axial antiferromagnetic spin-1/2 XXZ chain in a longitudinal magnetic field.
arXiv:1702.01854 (2017).

\bibitem{KnolleThesis}
J. Knolle,
\emph{Dynamics of a Quantum Spin Liquid},
(Springer, 2016), Chapter~5.

\end{thebibliography}
\end{document}